\documentclass[twocolumn,showpacs,preprintnumbers,amsmath,amssymb]{revtex4}
\usepackage{tabularx,graphicx}\begin{document}
\newcommand{\beq}{\begin{equation}}
\newcommand{\eeq}{\end{equation}}
\newcommand{\beqn}{\begin{eqnarray}}
\newcommand{\eeqn}{\end{eqnarray}}
\newcommand{\bmath}{\begin{subequations}}
\newcommand{\emath}{\end{subequations}}
\title{Do superconductors violate Lenz's law? Body rotation under field cooling and theoretical implications}
\author{J. E. Hirsch }
\address{Department of Physics, University of California, San Diego\\
La Jolla, CA 92093-0319}

\begin{abstract} 
When a magnetic field is turned on, a superconducting body acquires an angular momentum in direction opposite to the applied field. This
gyromagnetic effect has been established experimentally and is understood theoretically. However, the corresponding situation when a superconductor
is cooled in a pre-existent field has not been examined. We argue that the conventional theory of superconductivity does not provide a prediction
for the outcome of that experiment that does not violate fundamental laws of physics, either  Lenz's law  or
conservation of angular momentum. The theory of hole superconductivity   predicts an
outcome of this experiment consistent with the laws of physics.  \end{abstract}
\maketitle

\begin{figure}
\resizebox{7.5cm}{!}{\includegraphics[width=7cm]{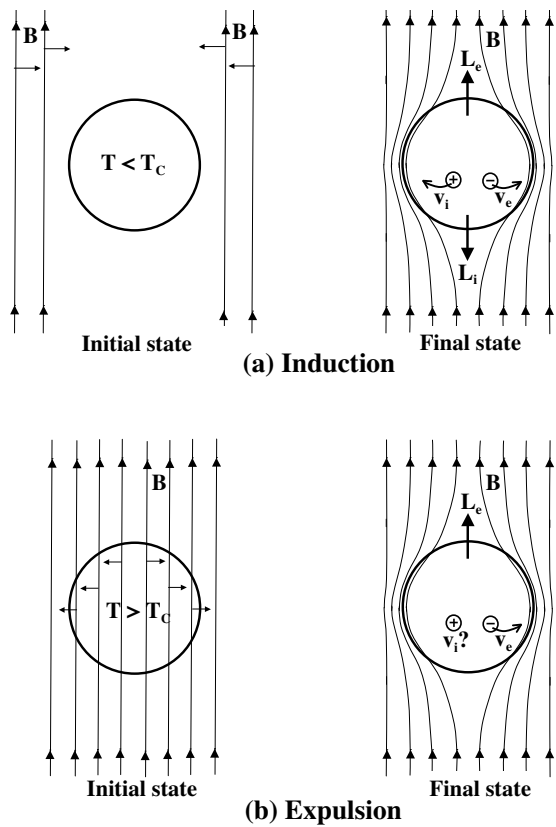}}
\caption{ (a) When a magnetic field is applied to a superconductor, the changing magnetic field induces a supercurrent of electrons with velocity $\vec{v}_e$ in the direction shown and angular
momentum $\vec{L}_e$ parallel to the field, and the positive ions (and the body) acquire 
velocity $\vec{v}_i$ and angular momentum $\vec{L}_i$ antiparallel to the field,
which has been measured experimentally. (b) When a normal metal is cooled in the presence of a magnetic field,
the supercurrent that develops is the same as in case (a), with angular momentum parallel to the field. The motion of the
ions and hence the body is unknown.
}
\label{figure1}
\end{figure} 
\section{introduction}
An electron circulating in a closed orbit generates a magnetic moment $\vec{\mu}$ related to its angular momentum $\vec{l}$ by
\beq
\vec{\mu}=\frac{e}{2m_ec}\vec{l}
\eeq
with $e$ ($<0$) and $m_e$ the electrons' charge and mass. When a magnetic field is applied to a superconducting body, it develops a
magnetization $\vec{m}$ to screen the magnetic field in its interior\cite{london}. It has been verified experimentally by
Kikoin and Gubar\cite{kikoin}, Pry et al\cite{pry} and Doll\cite{doll} that the body acquires an angular momentum $\vec{L}_i$ related to
its magnetization $\vec{m}$ by
\beq
\vec{L}_i=-\frac{2m_ec}{e}\vec{m}
\eeq
This is understood according to Eq. (1) if the carriers of the supercurrent are assumed to be (bare\cite{eha}) electrons carrying total angular momentum $\vec{L}_e$, with
\beq
\vec{L}_e+\vec{L}_i=0
\eeq
since the application of the external magnetic field will not change the total angular momentum of a 
homogeneous charge-neutral system. The finite angular
momentum acquired by both electrons and ions can be understood as arising from Faraday induction\cite{meissner}. Since the induced magnetization
$\vec{m}$ points antiparallel to the field, so does the ionic (positive charge) angular momentum (Eq. (2)), which is manifested in the rotation
of the body detected experimentally. The situation is schematically depicted in Fig. 1(a).

Consider now a type I superconductor that is cooled through its critical temperature in the presence of an external magnetic field (Fig. 1(b)). The magnetic field
is expelled from the interior (Meissner effect) and the final state {\it for the electrons} is the same as in the case discussed above. Is that
also true for the state of the ions? This question has not been addressed either theoretically or experimentally before, however it  is implicitly expected that the answer is affirmative. We argue here that this expectation is  incorrect.

\section{the puzzle}
 
First we need to ask: what causes the change in the total electronic mechanical angular momentum 
\beq
\vec{L}_e=m_e\sum_\alpha \vec{r}_\alpha  \times \vec{v}_\alpha 
\eeq
($\alpha$ labels individual electrons)
from its initial zero value in the normal state to its finite value carried by the Meissner current?\cite{lorentz} 
In a system of interacting particles, the time dependence of the total mechanical 
angular momentum is given by\cite{marion}
\beq
\frac{d\vec{L}_e}{dt}=\sum_\alpha \vec{r}_\alpha  \times \vec{F}_\alpha^{(e)}+\sum_{\alpha\neq\beta} \vec{r}_\alpha  \times \vec{f}_{\alpha \beta}
\eeq
where the first term is the external torque, due to external forces $\vec{F}_\alpha^{(e)}$, and the second term gives the internal torque due to interparticle forces $\vec{f}_{\alpha \beta}$. Under the reasonable assumption that the internal forces between electrons 
are central, the second term vanishes and the
change in angular momentum is given by
\beq
\frac{d\vec{L}_e}{dt}=\sum_\alpha \vec{r}_\alpha  \times \vec{F}_\alpha^{(e)}\equiv \vec{N}_e.
\eeq
with $\vec{N}_e$ the net external torque.

However, in the conventional theory of superconductivity there is no net external torque in the presence of an applied magnetic field. The
velocity of the electrons in the normal state points randomly in all directions, hence there is no net magnetic Lorentz force; and since we are
not externally changing the  magnetic field there is no induced azimuthal electric field as in the first case discussed above. This would imply  that
\beq
\frac{d\vec{L}_e}{dt}=0
\eeq
at all times, 
and since $\vec{L}_e$ is zero in the normal state, it can never reach the finite value given by Eq. (4) when the system is cooled into the
superconducting state. We are thus faced with a conundrum.

One may attempt to circumvent this difficulty by postulating that somehow the ions acquire an angular momentum in opposite direction
$\vec{L}_i$, so that Eq. (3) holds at all times  (above $T_c$, $\vec{L}_e=\vec{L}_i=0$), consistent with the absence of a net external torque throughout the process. However we argue that such assumption is untenable.

\begin{figure}
\resizebox{7.5cm}{!}{\includegraphics[width=7cm]{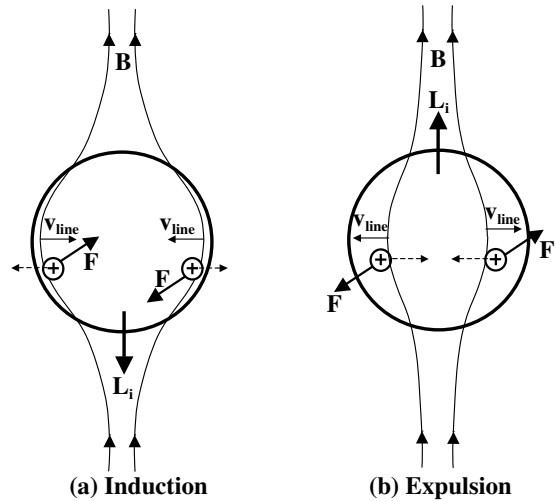}}
\caption{ (a) When a magnetic field is applied to a superconductor, the magnetic field lines move from the
outside into a surface layer of thickness given by the London penetration depth. The Lorentz force on the positive ions
caused by the magnetic
field lines moving in gives a torque in direction $antiparallel$ to the magnetic field.
(b) When the superconductor is cooled in a field magnetic field lines are expelled, and the Lorentz force on the
positive ions gives a torque   in direction $parallel$ to the magnetic field.
}
\label{figure2}
\end{figure}

Indeed, for the electrons, one might invoke some mysterious "quantum force"\cite{nikulov}, inherent in the superconductivity phenomenon,
 that drives the electrons to acquire the
final state with azimuthal velocity needed for the Meissner current. However,   the ions   are essentially classical objects (if pressed we would say that we have in mind an electronic
superconductivity mechanism where the phonons play absolutely no role). 
The force
that makes the ions rotate in the first case discussed above can be understood as the magnetic Lorentz force
\beq
\vec{F}=\frac{q}{c}\vec{v}\times\vec{B}
\eeq
that acts on the positive ions within a London penetration depth of the surface when the magnetic field lines penetrate that
region. In Eq. (8), $\vec{v}$ is the $relative$ speed of the ion with respect to the magnetic field line (dashed arrows in Fig. 2).
As depicted in Fig. 2(a), the torque is opposite to the applied magnetic field in this case, as expected.

However, in the second case under consideration, as the temperature is lowered below $T_c$  the magnetic field lines are 
pushed $out$ from the interior of the superconductor towards the surface. The resulting Lorentz force on positive ions
from such magnetic field line motion will   give rise to a torque {\it in the same direction} as the applied magnetic field,
as depicted in Fig. 2(b).

In other words: in the first case, the net flux of magnetic field in the interior of the sphere increased, in the second case it decreased.
The resulting induced azimuthal electric field from Faraday's law points in opposite directions, hence it pushes the
positive ions in   opposite directions. When the metal is cooled into the superconducting state the ions, following Lenz's law, should be trying to restore the magnetic flux that is being expelled from the interior, not the
opposite.
However, if the ions move in the direction dictated by Lenz's law, the ionic angular
momentum will point parallel to $\vec{B}$ and hence parallel to the electronic
angular momentum, and the sum of electronic and ionic angular momentum
will be non-zero.

 \section{  angular momentum conservation}

Under what conditions may the mechanical angular momentum of a system not be conserved? As argued earlier,
there has to be a net external torque.    Consider the torque that results from the Lorentz force on an electron in the presence of a magnetic field $\vec{B}$:
\beq
\frac{d\vec{L}_\alpha}{dt}=\frac{e}{c}  \vec{r}_\alpha  \times (\vec{v}_\alpha \times \vec{B})
\eeq
In cylindrical coordinates
\beq
\vec{v}_\alpha \equiv v_{\theta,\alpha} \hat{\theta} +v_{r,\alpha} \hat{r} + v_{z,\alpha} \hat{z}
\eeq
and for $\vec{B}=B\hat{z}$ we have
\beq
\frac{d\vec{L}_\alpha}{dt}=-\frac{e}{c} B [r_\alpha v_{r,\alpha} \hat{z} -z_\alpha v_{\theta,\alpha} \hat{\theta} +z_\alpha v_{r,\alpha} \hat{r})
\eeq
We can ignore the second and third terms that will average to zero by symmetry across the $z=0$ plane, and conclude that the rate of change of the 
total electronic angular momentum is
\beq
\frac{d\vec{L}_e}{dt}=\sum_\alpha \frac{d\vec{L}_\alpha}{dt}=-\frac{e}{c} [\sum_\alpha r_\alpha v_{r,\alpha}]\vec{B}
\eeq
The quantity in brackets gives the {\it radial flux of electrons}; recalling that $e<0$, Eq. (12) implies that to generate a net electronic 
mechanical angular 
momentum in the direction of $\vec{B}$ {\it requires a net outgoing flux of electrons in direction perpendicular to the magnetic field as the system enters the superconducting state.}

The conventional theory of superconductivity assumes that charge remains uniformly distributed in the superconducting state, hence
no net external torque from the magnetic field can be exerted. 
Hence the conventional theory implies that either Lenz's law is violated by the ionic motion $or$ that the conservation of
mechanical angular momentum in the absence of external torque is violated as a normal metal is cooled into the superconducting
state in the presence of a magnetic field. Clearly, neither choice is satisfactory.

Note that   the expulsion of magnetic field from an $insulating$ system would not be expected to result in rotation of the charges.
This is because in the insulator positive and negative charges are bound to each other, and the force exerted on them by the
induced electric field is in opposite direction. In the superconductor we may assume similarly that no rotation occurs in the interior if
the negative superfluid is "bound" to the positive  ions. However, within a London penetration depth of the surface the superfluid moves freely to generate the Meissner field,
and in that region one would expect the ions to respond to the induced electric field, just like in the case where magnetic field
lines enter from the outside (Fig. 1a), but in the opposite direction.

\section {Charge expulsion}
The theory of hole superconductivity\cite{holesc,holesc2} may offer a way out of this conundrum, because it
 predicts that electrons are expelled from the interior towards the surface of the body when a metal enters the
 superconducting state\cite{charge}. According to Eq. (12), this allows for the mechanical angular momentum of the system to change.
 
 We consider for simplicity a cylindrical geometry. Assume an electron is expelled from initial radius $r_1$ to final radius $r_2$.
 Responding to the Lorentz force it  will acquire an angular momentum in the direction of $\vec{B}$, given by
 \beq
 \vec{l}=\frac{e}{2c}(r_1^2-r_2^2)\vec{B}
 \eeq
 corresponding to the electron acquiring a tangential velocity
 \beq
 v_\phi=-\frac{e}{2 \pi m_e c}\frac{\Delta \phi}{r_2}
 \eeq
 with $\Delta \phi$ the change in the magnetic flux enclosed by the orbit
 \beq
 \Delta \phi=\pi(r_2^2-r_1^2)B
 \eeq
 This result is easily obtained by integration of the equation of motion, or by conservation of
 the canonical angular momentum $\vec{r}\times \vec{p}$, with $\vec{p}=m_e  \vec{v}+(e/c)\vec{A}$,
 ${A}(r)=\phi(r)/2\pi r$ the magnetic vector potential. Above $T_c$ the tangential velocity
 of electrons is zero on the average. If electrons move outward as the system enters the
 superconducting state, the expelled electrons will acquire a net angular velocity that will give rise
 to a Meissner current, and   associated mechanical angular momentum parallel to the applied
 magnetic field.
 
 For a cylinder of radius $R$ and height $h$, the total mechanical 
 angular momentum acquired by $N$ electrons expelled uniformly from the interior to the
 surface is
 \beq
 \vec{L}_e=-\frac{NeR^2}{4c}\vec{B}
 \eeq
 leaving a uniform positive charge distribution
 \beq
 \rho_0=-\frac{Ne}{\pi R^2 h}
 \eeq
 in the interior. Using the relation between electronic mechanical angular momentum and
 magnetic moment
 \beq
 \vec{L}_e=\frac{2m_ec}{e}\vec{m}
 \eeq
 we can relate $\vec{L}_e$ to the   magnetic field induced by the surface
 current  $\vec{B}_{ind}=4\pi\vec{M}$,  $\vec{M}=\vec{m}
 /(\pi R^2 h)$ as
 \beq
 \vec{L}_e=\frac{m_ec}{2e}R^2h \vec{B}_{ind}
 \eeq
 Now in Eq. (16), $\vec{B}$ is the $net$ magnetic field, i.e. the difference between the applied
 magnetic field ($\equiv \vec{B}_0$) and the induced field opposing it
 \beq
 \vec{B}=\vec{B}_0-\vec{B}_{ind}
 \eeq
 and from Eqs. (16), (19) and (20) we obtain
 \beq
 \vec{B}_{ind}=-\frac{\vec{B}_0}{1+\frac{2h}{Nr_e}}
 \eeq
 with $r_e$ the classical electron radius, $r_e=e^2/m_e c^2$.
 
 The mechanical angular momentum acquired by the expelled electrons is 
 compensated by angular momentum stored in the electromagnetic
 field
 \beq
 \vec{L}_{field}=\frac{1}{4\pi c}\int d^3r \vec{r}\times(\vec{E}\times\vec{B})
 \eeq
This is an example of the "Feynman Paradox" discussed in ref. \cite{feynman}. The electric field originating from the resulting positive charge distribution
 in the interior Eq. (17)   is
 \beq
 E(r)=\frac{2Ne}{h}\frac{r}{R^2}
 \eeq
 and Eqs. (22) and (23) yield
 \beq
 \vec{L}_{field}=-\vec{L}_e
 \eeq
 with $\vec{L}_e$ the mechanical angular momentum Eq. (16). Consequently, the sum of the
 mechanical angular momentum of the electrons and the angular momentum of the electromagnetic field remain zero throughout
 the process, with no contribution from ionic angular momentum.
 
 Indeed, within this scenario, we would not expect the ions to acquire any angular momentum. Here, the Meissner
 current is carried by the excess electrons expelled from the interior rather than by the
 electrons that already reside in the surface layer for $T>T_c$. The outgoing magnetic field lines
sweep through  the charge-neutral surface layer and as a consequence the net momentum transferred to electrons
 plus ions is zero. In other words, within this scenario the ions are not violating Lenz's law by not
 responding to the induced azimuthal electric field.

 \section{discussion}
 
 Eq. (21) shows that the expelled electrons will screen the applied magnetic
 field $\vec{B}_0$. In a sense, as the electrons flow from the interior towards the surface they
 drag with them the magnetic field lines. Such a behavior is expected in
 a perfectly conducting classical plasma\cite{plasma}. For example, for a 
 $99\%$ suppression of the interior magnetic field  the number of expelled electrons
 required from Eq. (21) is
 \beq
 \frac{N}{h}=\frac{2}{r_e}\times 99
 \eeq
 and the resulting positive charge density in the interior is
 \beq
 \rho_0=\frac{198 |e|}{\pi R^2 r_e}
 \eeq
 For example, for a sample of $1cm$ radius
 \beq
 \rho_0/|e|=2.24\times 10^{-10} \frac{electrons}{\AA^3}
 \eeq
 which shows that the number of expelled electrons required for an almost complete Meissner effect
 is a small fraction of the total number of electrons.
 
 In this classical model a full Meissner effect is never achieved. Nevertheless, it suggests that charge
 expulsion is an essential ingredient of the Meissner effect in superconductors. Furthermore,
 it provides a natural explanation for how superconductors can expel magnetic fields without 
 violating Lenz's law and angular momentum conservation, which is not explained by the conventional theory.
 
  \begin{figure}
\resizebox{4.5cm}{!}{\includegraphics[width=7cm]{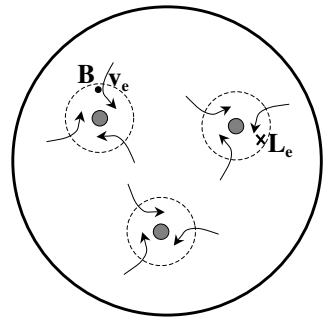}}
\caption{Type II superconductor cooled in a magnetic field (pointing out of the paper).  The shaded circles depict the vortex cores, the dashed circles indicate the boundary of the vortex currents.  Electrons flowing from the superconducting regions into the dashed circles will be deflected
by the Lorentz force of the applied magnetic field in the direction shown, resulting in creation of
mechanical angular momentum and supercurrent flow in the flux tubes in the direction required to enhance
the magnetic field inside the flux tubes. }
\label{figure2}
\end{figure} 

  \begin{figure}
\resizebox{4.5cm}{!}{\includegraphics[width=7cm]{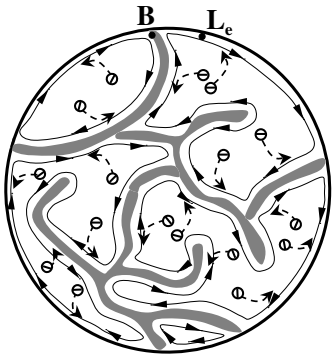}}
\caption{Type I superconductor cooled into an intermediate state with  laminar patterns. 
The shaded regions indicate normal
regions with magnetic flux through them. The full lines with arrows denote the path
of the supercurrents in the final
state.
As the system is cooled from above to below $T_c$, electrons flow from the superconducting
regions towards the shaded regions and towards the outer surface of the disk and   are deflected
by the Lorentz force, building up the currents to sustain the random resulting pattern. 
The electron trajectories are indicated by the dashed lines with arrows. }
\label{figure3}
\end{figure} 

 Finally, the charge expulsion scenario can also explain qualitatively the current patterns that develop in type II
 superconductors cooled in the presence of a magnetic field, and in type I superconductors entering the
 intermediate state. In Fig. 3, the electrons are expelled from the superconducting region of a type II material
 towards the normal vortex cores, and are deflected by the Lorenz force as shown to give rise to the 
 required vortex current around the flux tubes. In Fig. 4, we show a typical laminar structure
 characteristic of the intermediate state. Electrons expelled towards the normal regions and towards the
 outer surface of the body will be deflected by the Lorentz force in the direction required to create the current
 patterns to sustain the intermediate state structure. In both cases the mechanical 
 angular momentum acquired by the
 supercurrents will be compensated by angular momentum stored in the electromagnetic field,
 as in the case discussed earlier, and there will be excess negative charge at the
 normal-superconducting phase boundary regions and at the outer surfaces,
 and excess positive charge deep in the superconducting regions where no currents flow.

 An experiment measuring the rotation of a type I superconducting body cooled in an external magnetic field
 may be able to provide direct experimental evidence for the physics discussed here. If the body does not rotate
 when the magnetic field is expelled,   or if it acquires any angular momentum that does not satisfy 
 $\vec{L}_i=-\vec{L}_e$, it will indicate that angular momentum has been stored in the electromagnetic
 field, which   requires the existence of an electric field and associated macroscopic charge
 inhomogeneity, as predicted by our model.
  
  \acknowledgements
  The author is grateful to Ivan Schuller for calling "Feynman's paradox" to his attention, to Tom O'Neil for
  very helpful discussions, and to a thorough anonymous referee for constructive comments.
 
\end{document}